\documentclass[epj]{webofc}
\usepackage[utf8]{inputenc}
\usepackage[varg]{txfonts}   % Web of Conferences font
\usepackage{booktabs}
\usepackage{xcolor}
\definecolor{darkred}{rgb}{0.4,0.0,0.0}
\definecolor{darkgreen}{rgb}{0.0,0.4,0.0}
\definecolor{darkblue}{rgb}{0.0,0.0,0.4}
\usepackage[bookmarks,linktocpage,colorlinks,
    linkcolor = darkred,
    urlcolor  = darkblue,
    citecolor = darkgreen]{hyperref}
\usepackage{subfigure}
\wocname{EPJ Web of Conferences}
\woctitle{Lattice2017}
\usepackage{tikz}
\usepackage{tikzscale}
\usetikzlibrary{arrows,shapes,decorations.markings,tikzmark,patterns,calc,plotmarks,backgrounds}

\newcommand*{\BUILDFIGS}{}

\begin{document}
%%%%%%%%%%%%%%%%%%%%%%%%%%%%%%%%%%%%%%%%%%%%%%%%%%%%%%%%%%%%%%%%%%%%%%%%%%%%%
%
\selectlanguage{english}
%----------------------------------------------------------------------------
\title{%
Multi-hadron spectroscopy in a large physical volume
}
%----------------------------------------------------------------------------
\author{%
\firstname{John} \lastname{Bulava}\inst{1} \and
\firstname{Ben} \lastname{H\"{o}rz}\inst{2}\fnsep\thanks{Speaker, \email{hoerz@uni-mainz.de}} \and
\firstname{Colin}  \lastname{Morningstar}\inst{3}
% etc.
}
%----------------------------------------------------------------------------
\institute{%
Dept. of Mathematics and Computer Science and CP3-Origins, University of Southern Denmark, Campusvej 55, 5230 Odense M, Denmark
\and
PRISMA Cluster of Excellence and Institute for Nuclear Physics, University of Mainz, Johann Joachim Becher-Weg 45, 55099 Mainz, Germany
\and
Department of Physics, Carnegie Mellon University, Pittsburgh, PA 15213, USA
}
%----------------------------------------------------------------------------
\abstract{%
We demonstrate the efficacy of the stochastic LapH method to treat all-to-all quark propagation on a $N_\mathrm{f} = 2+1$ CLS ensemble with large linear spatial extent $L = 5.5 \, \mathrm{fm}$, allowing us to obtain the benchmark elastic isovector $p$-wave pion-pion scattering amplitude to good precision already on a relatively small number of gauge configurations. These results hold promise for multi-hadron spectroscopy at close-to-physical pion mass with exponential finite-volume effects under control.
}
%----------------------------------------------------------------------------
\maketitle
%----------------------------------------------------------------------------
\section{Introduction}\label{intro}

Hadron spectroscopy has a long-standing history in lattice QCD.
State-of-the-art calculations employ multi-hadron interpolating operators in addition to the traditional single-meson and single-baryon operators to accurately determine the spectrum of QCD in a finite volume, which can then be used to constrain infinite-volume physics via the L\"{u}scher formalism and its extensions \cite{Luscher:1990ux,Luscher:1991cf,Rummukainen:1995vs}.\footnote{For a recent review of the various extensions and applications of the L\"{u}scher formalism, see Ref.~\cite{Briceno:2017max}.}
In a finite volume with periodic boundary conditions in the spatial directions, information about infinite-volume scattering amplitudes involving QCD-stable particles is encoded in the energy shifts between eigenstates of the interacting theory and the corresponding noninteracting energies.
The leading term in the s-wave threshold expansion,
\begin{align}
	\Delta E = \frac{4\pi a_0}{m_\pi L^3} + \mathcal{O}\left ( L^{-4} \right ),
	\label{eqn:thresholdexpansion}
\end{align}
with $a_0$ the scattering length, makes it clear that the signal one has to discern vanishes with the physical spatial volume $L^3$, so that large physical volumes might appear unappealing at first.

However there is a compelling case in favor of large physical volumes: Firstly, corrections to Eq.~\eqref{eqn:thresholdexpansion} which are exponentially suppressed in $L$ are assumed to be negligible throughout the derivation of the L\"{u}scher formalism.
Large physical volumes are hence mandatory for the relationship between finite-volume and infinite-volume observables to remain applicable as the pion mass is lowered towards its physical value.
Secondly, as we will show in this contribution, employing large physical volumes is also beneficial since the number of finite-volume eigenstates in a given kinematical range increases with the spatial volume, so that many data points can be used to constrain the energy dependence of scattering amplitudes.

Computing the required correlation functions of multi-hadron operators, i.e. operators containing more than one hadron interpolator each projected to definite momentum individually \cite{Morningstar:2013bda}, is technically challenging and necessitates the use of all-to-all quark propagators.
Distillation \cite{Peardon:2009gh} has been tremendously successful as a method to treat all-to-all quark propagation, facilitating the computation of correlation functions to a precision sufficient for L\"{u}scher analyses (see e.g. Refs.~\cite{Lang:2011mn,Dudek:2012gj,Dudek:2012xn} for early results using distillation, and Ref.~\cite{Briceno:2017max} for a review of more recent results).
In large physical volumes however distillation becomes increasingly expensive.
This volume scaling of computational cost is ameliorated at the expense of a slight decrease in precision in the stochastic LapH method \cite{Morningstar:2011ka} by introducing a diluted \cite{Foley:2005ac} stochastic estimator in the distillation subspace.

In this contribution, we show that the stochastic LapH method yields sufficiently precise correlation functions to extract the Breit-Wigner parameters of the $\rho$-resonance on an ensemble with large spatial extent $L = 5.5\,\mathrm{fm}$ at a fraction of the cost of exact distillation.
Along the way we study the interplay between dilution and the number of noise sources with regard to attainable precision, as well as the volume dependence of dilution efficiency.
\hyperref[sec:alltoall]{Section~\ref*{sec:alltoall}} gives a brief overview of the stochastic LapH method to settle notation, and results are presented in \autoref{sec:results}.

%----------------------------------------------------------------------------
\section{The stochastic LapH method}\label{sec:alltoall}

In the stochastic LapH method \cite{Morningstar:2011ka}, the inverse of the Dirac matrix $M^{-1}$ is estimated by the outer product
\begin{align}
	M^{-1} \approx \sum_{r=1}^{N_\eta} \sum_a \varphi^{r,[a]} \varrho^{r,[a] \dagger}
	\label{eqn:stochest}
\end{align}
of a diluted random source vector $\varrho^{r,[a]}$ with dilution index $a$, noise index $r$, and the corresponding sink vector $\varphi^{r,[a]}$ given by
\begin{align}
	\varrho^{r,[a]} = V_s P^{[a]} \rho^r, \qquad \varphi^{r,[a]} = \mathcal{S} M^{-1} \varrho^{r,[a]},
	\label{eqn:stochlaph}
\end{align}
with orthogonal dilution projectors $P^{[a]}$ acting in time, spin and Laplacian eigenvector space and $\rho^r$ a random vector in that space with each component filled with $\mathbb{Z}_4$ noise.
The smearing matrix $\mathcal{S} = V_s V_s^\dagger$ projects into the distillation or LapH subspace spanned by the $N_\mathrm{ev}$ low-lying eigenvectors of the three-dimensional gauge covariant Laplacian on each time slice of the lattice.
On the level of interpolating operators, the application of $\mathcal{S}$ amounts to a spatial smearing of the quark fields with a Gaussian profile whose width is determined by the number of retained eigenvectors \cite{Peardon:2009gh}.\footnote{The smearing radius -- or equivalently the highest retained eigenvalue of the Laplacian -- should be tuned such that high-frequency fluctuations in the quark fields are sufficiently suppressed while still avoiding oversmearing. A study of the sensitivity of spectral quantities was presented at last year's Lattice conference \cite{Woss:2016tys}.}
In order to keep the smearing radius constant, the number of eigenvectors to retain is empirically found to scale linearly with the physical volume, $N_\mathrm{ev} \sim L^3$.
This scaling is at the heart of the increase of computational cost with the volume in exact distillation, and is overcome in the stochastic LapH method by estimating the Dirac matrix inverse in the distillation subspace stochastically.

The rationale behind our use of a stochastic estimator in the LapH subspace is that the achievable precision of a lattice QCD observable is ultimately limited by the finite sampling of the path integral, and it is hence sufficient to estimate the quark propagator to a precision comparable with the final \emph{gauge noise}.
To this end, dilution \cite{Foley:2005ac} facilitates stochastic estimates of $M^{-1}$ with significantly reduced variance.
Following the nomenclature of Ref.~\cite{Morningstar:2011ka} we use full dilution in spin (SF) and interlace dilution in Laplacian eigenvector space so that, for a given dilution index $a$, the source is built from eigenvectors which are interleaved with a distance of $N_\mathrm{dil}$ in the eigenvector index (LI$N_\mathrm{dil}$ for short).
The exact distillation result is recovered by setting $N_\mathrm{dil} = N_\mathrm{ev}$ which we refer to as the gauge noise limit.
In practice the goal is to get sufficiently close to the distillation result with a moderate number of required inversions $N_\mathrm{inv} \propto 4 N_\eta N_\mathrm{dil}$ per configuration.
This interplay between the level of dilution and the number of noise vectors used to estimate the quark lines in the computation of correlation functions  is the subject of \autoref{subsec:dilutioneff}.

\section{Results}\label{sec:results}

\subsection{Ensemble details}
\begin{table}[thb]
  \small
  \centering
  \caption{Ensembles used in this work, which have been generated through the CLS effort. Details about the simulations can be found in Ref.~\cite{Bruno:2014jqa} and the scale setting is discussed in Ref.~\cite{Bruno:2016plf}. Both ensembles share the same bare parameters and differ only in their volume.}
  \begin{tabular}{c c c c c c}\toprule
   & $a \, [\mathrm{fm}]$ & $m_\pi \, [\mathrm{MeV}]$ & $N_\mathrm{cfg}$ & $(L/a)^3 \times T/a$ & $m_\pi L$  \\\midrule
  D101 & $0.086$ & $220$ & $132$ & $64^3 \times 128$ & $6.2$ \\
  C101 &  & & & $48^3 \times 128$ & $4.7$ \\
  \bottomrule
  \end{tabular}
  \label{tab:ensembles}
\end{table}
The results presented here are part of our ongoing effort to obtain a precise determination of the properties of the $\rho$-resonance \cite{Bulava:2015qjz} with all systematic uncertainties under control, based on the $N_\mathrm{f} = 2+1$ ensembles generated through the Coordinated Lattice Simulations (CLS) effort \cite{Bruno:2014jqa,Bruno:2016plf} (see \autoref{tab:ensembles} for some relevant parameters of the two ensembles used here).

\begin{table}[thb]
  \small
  \centering
  \caption{Stochastic LapH parameters on the two ensembles employed in this work. The $N_\mathrm{ev}$ eigenvectors of the three-dimensional Laplacian are computed after applying $n_\rho = 20$ steps of three-dimensional stout smearing \cite{Morningstar:2003gk} with $\rho = 0.1$ to the gauge links, tuned such that in physical units $n_\rho \rho = \mathrm{const}$ compared to previous studies \cite{Morningstar:2011ka,Bulava:2016mks}. Source time positions are chosen so as to avoid boundary effects from the open temporal boundaries. Naming conventions for quark line types and dilution schemes follow Ref.~\cite{Morningstar:2011ka}.}
  \begin{tabular}{c c c c c c}\toprule
    & $N_\mathrm{ev}$ & line type & dilution scheme & $t_0/a$ & $N_\eta$ \\\midrule
  D101 & $928$ & fixed & (TF,SF,LI16) & $24,44,64$ & $8$ \\
   & & relative & (TI8,SF,LI16) & -- & $3$ \\\midrule
  C101 & $392$ & fixed & (TF,SF,LI16) & $20,40$ & $6$ \\
   & & relative & (TI8,SF,LI16) & -- & $2$ \\
  \bottomrule
  \end{tabular}
  \label{tab:laphparameters}
\end{table}
The two ensembles differ only in their physical volume with all other parameters fixed, with the bigger one of the two boxes having a linear spatial extent of $L = 5.5 \, \mathrm{fm}$, which to the best of our knowledge is the biggest volume to be used to date for multi-hadron spectroscopy with LapH smearing.
The remaining LapH parameters are collected in \autoref{tab:laphparameters}, and the number of eigenvectors in the definition of the smearing is chosen such that the smearing radius is approximately the same on both ensembles.
While the D101 is the main ensemble used in this study, the C101 ensemble allows for a study of the volume (or equivalently $N_\mathrm{ev}$) dependence of dilution efficiency.
For the setup outlined in \autoref{tab:laphparameters}, the number of inversions required on each gauge configuration of the D101 ensemble is $N_\mathrm{inv} = 4 N_{t_0} N^\mathrm{fix}_\eta N_\mathrm{dil} + 32 N^\mathrm{rel}_\eta N_\mathrm{dil} = 3072$, to be compared to $N_\mathrm{inv} = 237,568$ inversions required for exact distillation on the central half of the lattice.
Thus, in the spirit of the discussion in Ref.~\cite{Morningstar:2011ka} the results using exact distillation would have to be almost an order of magnitude more precise to justify the additional computational effort.
\subsection{Dilution efficiency}\label{subsec:dilutioneff}
Various observables can in principle be used to assess the efficacy of the stochastic estimator with dilution.
The most natural choice would arguably be the difference between the exact quark propagator in the LapH subspace and our stochastic estimate \eqref{eqn:stochest} under some suitable matrix norm.
However since this metric requires knowledge of the distillation result and is thus prohibitively expensive on the D101 ensemble, we resort to more indirect but physically relevant observables and consider the correlation functions of a single $\rho$-meson and of a pion instead.\footnote{Another option would be to consider effective masses of the respective correlation functions, but the $\rho$-like interpolator has weak overlap with the {\lq}two-pion--like{\rq} ground state in the at-rest irreducible representation. In particular, using effective masses would thus complicate the comparison between the ensembles with different volumes.}
\begin{figure}[thb]
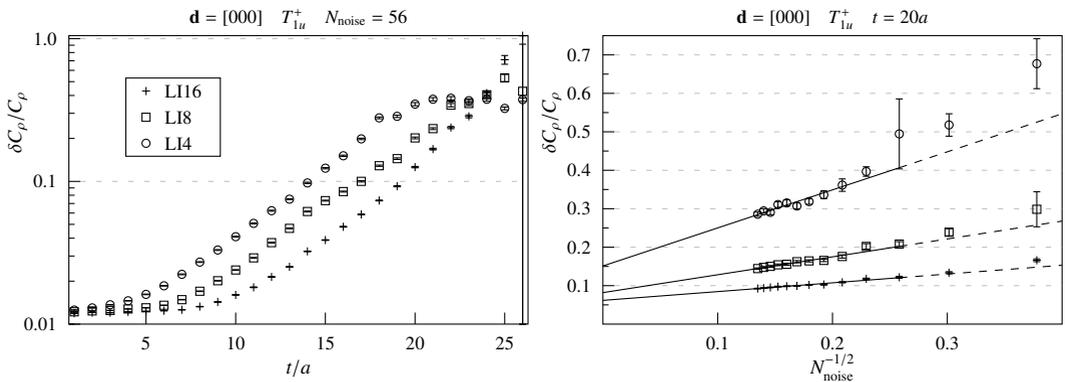

  \centering
  \ifdefined\BUILDFIGS
  \includegraphics[width=0.49\textwidth]{img/d101_rho_sn.tikz}
  \includegraphics[width=0.49\textwidth]{img/d101_rho_scaling.tikz}
  \fi
  \caption{\textit{Left:} Noise-to-signal ratio of the single-$\rho$ correlator as a function of time separation for three different levels of dilution. \textit{Right:} Scaling of single-$\rho$ correlator uncertainty at fixed time separation $t/a=20$ with the number of noise combinations $N_\mathrm{noise}$ for each dilution. The lines correspond to fits assuming Monte Carlo error scaling in the region of the solid line. In both plots the uncertainty of the correlator is estimated from bootstrap resampling in gauge configurations, and its error estimated from resampling in the available noise combinations used to evaluate the correlation function. Evidently this estimate of the error of the error gives only some indication and should be interpreted with care at large time separations.}
  \label{fig:rhoscaling}
\end{figure}
Those correlation functions feature only connected contributions and provide a clean testbed to assess the efficacy of eigenlevel dilution in the estimation of fixed lines, without the additional complications due to dilution in time used for relative lines.
For a given number of noise sources $N_\eta$ used in inversions, there are $N_\mathrm{noise} = N_\eta \left( N_\eta -1 \right)$ ways to estimate a single-meson correlation function.
In addition, if inversions have been performed with dilution scheme LI$N_\mathrm{dil}$, the data can be undiluted \emph{a posteriori} to obtain the results with LI$N_\mathrm{dil}/2$ dilution, thus allowing for a systematic study of the interplay of $N_\mathrm{noise}$ and $N_\mathrm{dil}$.

The noise-to-signal ratio of the \textit{single-site} \cite{Morningstar:2013bda} $\rho$ correlator as a function of time separation using the maximal number of noise combinations is shown in the left panel of \autoref{fig:rhoscaling}.
The signal-to-noise problem of the $\rho$ is clearly visible for each dilution scheme, its onset however can be pushed to later times if higher levels of dilution are afforded.

In the right panel of \autoref{fig:rhoscaling} the correlator uncertainty is plotted for a fixed time separation instead, showing its scaling with the number of noise combinations $N_\mathrm{noise}$ used in the computation of the correlation function.
The stochastic estimation of the quark propagator can be viewed as as an application of the Monte Carlo method inside the Monte Carlo sampling of the path integral, but it is not \emph{a priori} clear if a scaling window with the number of measurements exists, because the $N_\mathrm{noise}$ noise combinations are built from only $N_\eta$ random sources and thus not independent.
Nevertheless the data is well described by a linear fit in $N_\mathrm{noise}^{-1/2}$; as expected dilution reduces the variance of the stochastic estimator, which translates into a smaller slope with increasing dilution.
Furthermore, the extrapolated values of the correlator uncertainty, which yield an approximate determination of the gauge noise, are consistent within a factor of two to three between the different dilution schemes, and to the extent that the estimate of the gauge noise can be trusted, the LI$16$ curve yields a precision which is rather close even at finite $N_\mathrm{noise}$.\footnote{At smaller time separations, the extrapolation to the joint gauge noise limit works even better. We opted to show time separations relevant for spectroscopy and do not reproduce the other figures here due to space constraints.}

\begin{figure}[thb]
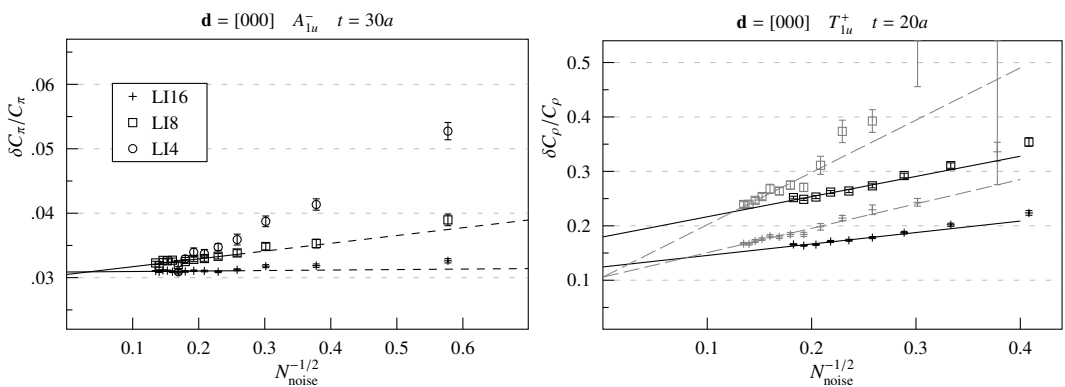

  \centering
  \ifdefined\BUILDFIGS
  \includegraphics[width=0.49\textwidth]{img/d101_pion_scaling.tikz}
  \includegraphics[width=0.49\textwidth]{img/c101_d101_rho_scaling.tikz}
  \fi
  \caption{\textit{Left:} Same as \autoref{fig:rhoscaling} but for the pion correlator at $t/a = 30$. Some LI$4$ data points at small $N_\mathrm{noise}^{-1/2}$ coinciding with the LI$16$ curve are not shown for better legibility. \textit{Right:} Comparison of the noise-dilution scaling of the $\rho$ correlator at $t/a=20$ on two different volumes, using $N_\mathrm{cfg}=132$ configurations of the D101 (dashed gray lines) and C101 (solid black lines) ensembles. Only LI$8$ and LI$16$ are shown, and only two out of three source times were used on the D101 ensemble for a fair comparison.}
  \label{fig:piscaling}
\end{figure}
In contrast to the $\rho$ correlator, the correlation function of a pion at rest does not have a signal-to-noise problem, and as shown in \autoref{fig:piscaling} a high level of dilution appears to be somewhat less critical: while even with very little dilution the gauge-noise limit is reached if a sufficient number of noise combinations is used, there again is a notable difference in the slope of noise scaling between different levels of dilution.
In particular the noise scaling is almost flat for LI$16$ dilution, a feature that could be exploited to inform the tuning of $N_\mathrm{dil}$.

The last aspect to study is how dilution efficacy changes as the volume (and $N_\mathrm{ev}$) is varied.
In the right panel of \autoref{fig:piscaling}, the combined noise and dilution scaling on the C101 and D101 ensembles are compared.
In line with expectations, the variance of the stochastic estimation increases with the dimension of the LapH subspace at fixed dilution.
However, although the number of eigenvectors, and hence the dimension of the matrix to be inverted, changes by a factor of roughly $2.3$ between both ensembles, there is no catastrophic breakdown in attainable precision.

Interlace-16 dilution appears to be {\lq}optimal{\rq} in the sense that the gain in precision by increasing the dilution (hence computer time required for the inversions) from LI$8$ to LI$16$ is roughly similar to the expected gain from increasing the number of noise sources.
Increasing the level of dilution further would thus presumably be less efficient than increasing the number of noises used in the stochastic estimates.

\subsection{Benchmark: pion-pion scattering amplitude}
\begin{figure}[thb]
  \centering
  \ifdefined\BUILDFIGS
  \includegraphics[width=0.85\textwidth]{img/d101_box_plot.tikz}
  \fi
  \caption{Finite-volume spectrum on the D101 ensemble in various irreducible representations (irreps) relevant for the $\rho$-resonance. The total momentum squared in units of $\left( 2\pi/L \right)^2$ is written in parenthesis for each irrep. Each box corresponds to the center-of-mass energy of a finite-volume stationary state with its width indicating the statistical uncertainty. Also shown are overlaps of some of the interpolating operators onto the different states with numbers in parentheses indicating squared momenta of constituent hadrons. Fill patterns identify the different energy levels in each irrep and are consistent between the upper and lower part of the figure. Not all interpolators are shown.}
  \label{fig:d101spectrum}
\end{figure}
In the last section we examined in detail how much variance is superimposed on correlation functions in addition to the gauge noise through our use of a stochastic estimation of the quark propagator in the LapH subspace.
Ultimately our goal is to perform L\"{u}scher-style analyses of the finite-volume spectra determined using the stochastic LapH method, and we demonstrate in this section that the precision of correlation functions on the D101 ensemble is indeed sufficient to map out the resonance shape of the $\rho$-resonance in isovector vector pion-pion scattering.
Our analysis strategy to extract the finite-volume spectrum has been described previously \cite{Bulava:2016mks}, and in the subsequent L\"{u}scher analysis we consider only elastic pion-pion scattering in the $\ell=1$ partial wave as higher partial waves are expected to be negligible \cite{Dudek:2012xn,Morningstar:2017spu}.

In \autoref{fig:d101spectrum} we plot the finite-volume spectrum on the D101 ensemble in various irreducible representations relevant to the $\rho$-resonance.
Due to the moderately low pion mass and the large physical volume the ground state in each irreducible representation is created predominantly by an interpolator that mimics two pions with some combination of momenta, and mixing of several operators is evident in the creation of states in the energy region where an extra state is expected due to the presence of the $\rho$-resonance.
Even high up in the spectrum energy levels can be extracted reliably, but since it is currently unknown how to relate the finite-volume spectrum to infinite-volume scattering amplitudes above four-particle thresholds it would be of limited use in our context to extract additional states.

\begin{figure}[thb]
  \centering
  \ifdefined\BUILDFIGS
  \includegraphics[width=0.69\textwidth]{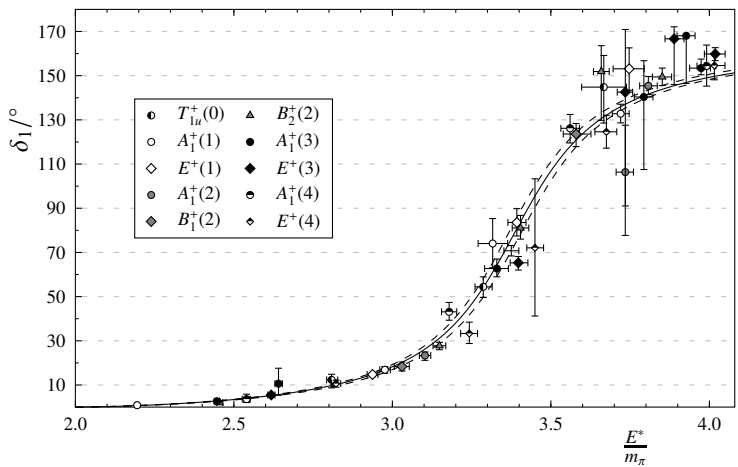}
  \fi
  \caption{Energy dependence of the phase shift describing the isovector $\ell=1$ pion-pion scattering amplitude. Every finite-volume energy level corresponds to a data point at the same center-of-mass energy. Resonance parameters are extracted from a Breit-Wigner fit to the data (solid line) with dashed curves indicating the statistical uncertainty. Two data points with error bars spanning the whole panel have been removed for legibility.}
  \label{fig:d101phase}
\end{figure}
As we restrict the L\"{u}scher analysis to the elastic region and neglect the effect of partial waves $\ell \ge 3$, every finite-volume energy level corresponding to a center-of-mass energy $E^* < 4 m_\pi$ completely determines the phase shift used to describe the pion-pion scattering amplitude at that center-of-mass energy.
We can thus map out the energy dependence of the pion-pion scattering amplitude with an unprecedented resolution (see \autoref{fig:d101phase}) and find the characteristic sharp increase through $90^\circ$ indicative of the $\rho$-resonance.
Due to the small width of the $\rho$-resonance, its shape is well described by a Breit-Wigner form used in a fit to determine the Breit-Wigner resonance parameters
\begin{align}
	m_r / m_\pi = 3.43(2), \qquad g_{\rho \pi \pi} = 6.05(11), \qquad \chi^2/\mathrm{d.o.f.} = 0.90.
	\label{eqn:rhobwpars}
\end{align}
The resonance mass $m_r$ and coupling constant $g_{\rho \pi\pi}$ are determined to percent-level precision already on the relatively small number of gauge configurations $N_\mathrm{cfg} = 132$ and using a low number of noise combinations $N_\mathrm{noise} = 9$ in the evaluation of single-meson correlation functions.
\section{Conclusions}
State-of-the-art lattice QCD spectroscopy calculations include multi-hadron interpolating operators in addition to single-hadron interpolators.
The stochastic LapH method facilitates the computation of the required correlation functions and the computational cost does not increase in proportion to the physical volume.
In this contribution we have taken the first steps to understand the scaling of attainable precision with dilution and the average over noise sources.
Remarkably the average over noise combinations is useful and consistent with the scaling expected from independent measurements even though the individual measurements are not independent.
Interlace-16 dilution in Laplacian eigenvector space is found to be a good choice up to the large spatial volume $V = \left( 5.5 \, \mathrm{fm} \right)^3$ considered here, and the attainable precision is sufficient to extract scattering amplitudes via L\"{u}scher-style analyses.

\textbf{Acknowledgements:} This work was supported by the U.S. National Science Foundation under award PHY-1613449.
The authors wish to acknowledge the DJEI/DES/SFI/HEA Irish Centre for High-End Computing (ICHEC) for the provision of computational facilities and support.
The authors gratefully acknowledge the Gauss Centre for Supercomputing e.V. (www.gauss-centre.eu) for funding this project by providing computing time through the John von Neumann Institute for Computing (NIC) on the GCS Supercomputer JUQUEEN at J\"{u}lich Supercomputing Centre (JSC).

%

%\clearpage
\bibliography{literature}

\begin{thebibliography}{18}

\bibitem{Luscher:1990ux}
M.~{L\"{u}scher}, Nucl. Phys. \textbf{B354}, 531 (1991)

\bibitem{Luscher:1991cf}
M.~{L\"{u}scher}, Nucl. Phys. \textbf{B364}, 237 (1991)

\bibitem{Rummukainen:1995vs}
K.~Rummukainen, S.A. Gottlieb, Nucl. Phys. \textbf{B450}, 397 (1995),
  \texttt{hep-lat/9503028}

\bibitem{Briceno:2017max}
R.A. {Brice\~{n}o}, J.J. Dudek, R.D. Young (2017), \texttt{1706.06223}

\bibitem{Morningstar:2013bda}
C.~Morningstar, J.~Bulava, B.~Fahy, J.~Foley, Y.C. Jhang, K.J. Juge,
  D.~Lenkner, C.H. Wong, Phys. Rev. \textbf{D88}, 014511 (2013),
  \texttt{1303.6816}

\bibitem{Peardon:2009gh}
M.~Peardon, J.~Bulava, J.~Foley, C.~Morningstar, J.~Dudek, R.G. Edwards,
  B.~Joo, H.W. Lin, D.G. Richards, K.J. Juge (Hadron Spectrum), Phys. Rev.
  \textbf{D80}, 054506 (2009), \texttt{0905.2160}

\bibitem{Lang:2011mn}
C.B. Lang, D.~Mohler, S.~Prelovsek, M.~Vidmar, Phys. Rev. \textbf{D84}, 054503
  (2011), [Erratum: Phys. Rev.D89,no.5,059903(2014)], \texttt{1105.5636}

\bibitem{Dudek:2012gj}
J.J. Dudek, R.G. Edwards, C.E. Thomas, Phys. Rev. \textbf{D86}, 034031 (2012),
  \texttt{1203.6041}

\bibitem{Dudek:2012xn}
J.J. Dudek, R.G. Edwards, C.E. Thomas (Hadron Spectrum), Phys.Rev.
  \textbf{D87}, 034505 (2013), \texttt{1212.0830}

\bibitem{Morningstar:2011ka}
C.~Morningstar, J.~Bulava, J.~Foley, K.J. Juge, D.~Lenkner, M.~Peardon, C.H.
  Wong, Phys. Rev. \textbf{D83}, 114505 (2011), \texttt{1104.3870}

\bibitem{Foley:2005ac}
J.~Foley, K.~Jimmy~Juge, A.~O'Cais, M.~Peardon, S.M. Ryan, J.I. Skullerud,
  Comput. Phys. Commun. \textbf{172}, 145 (2005), \texttt{hep-lat/0505023}

\bibitem{Woss:2016tys}
A.J. Woss, C.E. Thomas, PoS \textbf{LATTICE2016}, 134 (2016),
  \texttt{1612.05437}

\bibitem{Bruno:2014jqa}
M.~Bruno et~al., JHEP \textbf{02}, 043 (2015), \texttt{1411.3982}

\bibitem{Bruno:2016plf}
M.~Bruno, T.~Korzec, S.~Schaefer (2016), \texttt{1608.08900}

\bibitem{Bulava:2015qjz}
J.~Bulava, B.~{H\"{o}rz}, B.~Fahy, K.J. Juge, C.~Morningstar, C.H. Wong, PoS
  \textbf{LATTICE2015}, 069 (2016), \texttt{1511.02351}

\bibitem{Morningstar:2003gk}
C.~Morningstar, M.J. Peardon, Phys. Rev. \textbf{D69}, 054501 (2004),
  \texttt{hep-lat/0311018}

\bibitem{Bulava:2016mks}
J.~Bulava, B.~Fahy, B.~{H\"{o}rz}, K.J. Juge, C.~Morningstar, C.H. Wong, Nucl.
  Phys. \textbf{B910}, 842 (2016), \texttt{1604.05593}

\bibitem{Morningstar:2017spu}
C.~Morningstar, J.~Bulava, B.~Singha, R.~Brett, J.~Fallica, A.~Hanlon,
  B.~{H\"{o}rz}, Nucl. Phys. \textbf{B924}, 477 (2017), \texttt{1707.05817}

\end{thebibliography}

%%%%%%%%%%%%%%%%%%%%%%%%%%%%%%%%%%%%%%%%%%%%%%%%%%%%%%%%%%%%%%%%%%%%%%%%%%%%%
\end{document}